\documentclass[conference]{IEEEtran}
\usepackage{mathptmx} 
\usepackage{soul,color}
\soulregister\cite7
\soulregister\ref7
\soulregister\pageref7

\usepackage{fancyhdr}
\usepackage[normalem]{ulem}
\usepackage[hyphens]{url}
\usepackage[sort,nocompress]{cite}
\usepackage[final]{microtype}
\usepackage{flushend}
\usepackage{graphicx}
\usepackage{subcaption}
\usepackage{comment}
\usepackage[section]{placeins}
\usepackage[ruled,vlined,linesnumbered]{algorithm2e}
\usepackage[bookmarks=true,breaklinks=true,letterpaper=true,colorlinks,linkcolor=black,citecolor=blue,urlcolor=black]{hyperref}
\usepackage{tikz}
\usepackage{xcolor}
\usepackage{amsmath}
\usepackage[nodisplayskipstretch]{setspace}
\usepackage[normalem]{ulem}

\usepackage{caption}
\def\BibTeX{{\rm B\kern-.05em{\sc i\kern-.025em b}\kern-.08em
    T\kern-.1667em\lower.7ex\hbox{E}\kern-.125emX}}

\newcommand{\hoda}[1]{\textcolor{blue}{Hoda: \em #1 }}
\newcommand{\sankha}[1]{\textcolor{red}{Sankha: \em #1 }}
\newcommand{\nael}[1]{\textcolor{orange}{Nael: \em #1}}
\newcommand{\andres}[1]{\textcolor{green}{\em #1 }}

\renewcommand{\andres}[1]{}
\newcommand{\cut}[1]{}

\newcommand*\circled[1]{\tikz[baseline=(char.base)]{
            \node[shape=circle,fill,inner sep=2pt] (char) {\textcolor{white}{#1}};}}
            
\fancypagestyle{firstpage}{
  \fancyhf{}

  \fancyfoot[C]{\thepage}
}

\title{Leaky Buddies: Cross-Component Covert Channels on Integrated CPU-GPU Systems} 
\makeatletter
\newcommand{\linebreakand}{%
  \end{@IEEEauthorhalign}
  \hfill\mbox{}\par
  \mbox{}\hfill\begin{@IEEEauthorhalign}
}
\makeatother

\author{\IEEEauthorblockN{Sankha Baran Dutta}
\IEEEauthorblockA{\textit{Department of CSE} \\
\textit{University of California, Riverside}\\
Riverside, California, USA \\
sdutt004@ucr.edu}
\and
\IEEEauthorblockN{Hoda Naghibijouybari}
\IEEEauthorblockA{\textit{Department of Computer Science} \\
\textit{Binghamton University}\\
Binghamton, New York, USA \\
hnaghibi@binghamton.edu}
\and
\IEEEauthorblockN{Nael Abu-Ghazaleh}
\IEEEauthorblockA{\textit{Department of CSE} \\
\textit{University of California, Riverside}\\
Riverside, California, USA \\
nael@cs.ucr.edu}

 \linebreakand 
\IEEEauthorblockN{Andres Marquez}
\IEEEauthorblockA{\textit{Pacific Northwest National
Laboratory} \\
Richland, WA, USA\\
Andres.Marquez@pnnl.gov}
\and
\IEEEauthorblockN{Kevin Barker}
\IEEEauthorblockA{\textit{Pacific Northwest National
Laboratory} \\
Richland, WA, USA\\
Kevin.Barker@pnnl.gov}
}

\begin{document}

\maketitle
\thispagestyle{firstpage}
\pagestyle{plain}


\begin{abstract}
Graphics Processing Units (GPUs) are a ubiquitous component across the range of today's computing platforms, from phones and tablets, through personal computers, to high-end server class platforms.  
With the increasing importance of graphics and video workloads, recent processors are shipped with GPU devices that are integrated on the same chip.  
Integrated GPUs share some resources with the CPU and as a result, there is a potential for microarchitectural attacks from the GPU to the CPU or vice versa. We believe this type of attack, crossing the component boundary (GPU to CPU or vice versa) is novel, introducing unique challenges, but also providing the attacker with new capabilities that must be considered when we design defenses against microarchitectrual attacks in these environments. 
 Specifically, we consider the potential for covert channel attacks that arise either from shared microarchitectural components (such as caches) or through shared contention domains (e.g., shared buses).  We illustrate these two types of channels by developing two reliable covert channel attacks.  The first covert channel uses the shared LLC cache in Intel's integrated GPU architectures.  The second is a contention based channel targeting the ring bus connecting the CPU and GPU to the LLC.  Cross component channels introduce a number of new challenges that we had to overcome since they occur across heterogeneous components that use different computation models and are interconnected using asymmetric memory hierarchies.   We also exploit GPU parallelism to increase the bandwidth of the communication, even without relying on a common clock. The LLC based channel achieves a bandwidth of 120 kbps with a low error rate of 2\%, while the contention based channel delivers up to 400 kbps with a 0.8\% error rate.  

\end{abstract}

\section{Introduction} \label{sec:intro}

In recent years, micro-architectural covert and side channel attacks have been widely studied on modern CPUs, exploiting optimization techniques and structures to exfiltrate sensitive information. A preponderance of these studies exploit CPU structures examining channels through a variety of contention domains including caches~\cite{mehmet-2016,liu-15, wang-06,percival-05, fan_hpca2018}, branch predictors~\cite{evtyushkin-16-branch}, random number generators~\cite{evtyushkin-16-random}, and others~\cite{chen-ccHunter-2014,power-covert}.  Modern computing systems are increasingly heterogeneous, consisting  of  a  federation  of  the  CPU  with  GPUs, NPUs, other  specialized accelerators, as well as memory and storage components, using a rich interconnect. 
It  is  essential to  understand  how  micro-architectural  attacks manifest within such complex environments (i.e., beyond just the CPU).  

\cut{Graphics Processing Units (GPU) have become an integral part of  platforms, present from end-user devices to large scale high performance clusters and data centers. Although GPUs are primarily used for graphics rendering purposes, a broad range of general purpose computational workloads are also being accelerated by GPUs; for example, interfaces such as WebGL provide general purpose GPU acceleration to webpages through JavaScript~\cite{webgl}. The Top 500 supercomputer list is dominated with systems using GPUs~\cite{top-500}. 
Infrastructure-as-a-Service (IaaS)~\cite{amazon-aws,micro-azure} cloud platforms offer GPU instances for their clients.  GPUs come in two forms (1) discrete GPUs: a separate device, which is connected with the rest of the systems typically using a PCIe bus, which addresses a separate physical memory; (2)} 

 We consider a new type of attack which we term {\em cross-component} attacks.  In these attacks, the attacker resides on a component within a heterogeneous system (e.g., an accelerator) and launches an attack on a victim executing on another component (e.g., the CPU, or another accelerator).  We explore the principles of such attacks by exploring covert channel attacks within integrated GPU (iGPU) systems, where a GPU is built on the same die as the main CPUs, exploring covert channel attacks originating from the GPU to the CPU or vice versa. 
iGPUs are integrated with consumer class CPUs used in desktops and laptops, and also extensively used in portable electronic devices such as tablets and smart phones to provide graphics, compute, media, and display capabilities.   Moreover, iGPUs exemplify a trend to gradually increase the level  of  heterogeneity  in  modern  computing  systems,  as  further  scaling  of  fabrication  technologies  allows   formerly discrete components to become integrated parts of a system-on-chip, and provides for integration of specialized hardware accelerators for important workloads.  Understanding microarchitectural vulnerabilities in such environments is essential to the security of these widely used systems, as well as to illuminate potential threats to general heterogeneous computing systems.
 
Although covert channels have been demonstarted on a variety of CPU structures, as well as on discrete GPUs~\cite{Hoda-MICRO,Hoda-CCS, jiang-2016, jiang-2017}, we believe our attacks are significantly different from prior work because they operate across heterogeneous components.  Specifically, to the best of our knowledge, all prior demonstrated covert channels are symmetric, with both the sender and receiver being identical: typically threads or processes access a resource that they use to create contention.  In contrast, cross-component attacks occur between two entities that can have substantially different computational models, and that share asymmetric access to resources.  As a result, the attacks necessitate careful reverse engineering of asymmetric views of the resource from both side, and understanding of how contention occurs between them across a complex interconnected architecture.  Moreover, we believe this is the first attack demonstrated on heterogeneous environments, providing important insights into how this threat model manifests in such systems, and extend our understanding of the threat model and guide further research into defenses.  It is also important to note that successfully creating a covert channel establishes the presence of leakage and is a pre-requisite indicator of the potential presence of the more dangerous side channels, which we will explore in future.

\cut{In particular, we explore micro-architectural covert channels that cross from the CPU to the GPU or vice versa.} 
 An iGPU is tightly integrated on the same die with the CPU and shares resources such  as  the  last  level  cache  and  memory  subsystem  with it, in contrast to discrete GPUs which have a  dedicated  graphics memory.  This integration opens up the potential of new attacks that exploit the use of common resources to create interference between these components, leading to cross-component micro-architectural attacks. Specifically, we develop covert channels (secret communication channels that exploit contention) on integrated heterogeneous systems in which two malicious applications, located on two different components (CPU and iGPU) transfer secret information via shared hardware resources. 
In order to develop this new type of channels, we had to solve a number of new challenges relating to synchronization across heterogeneous components with frequency disparity, reconciling different computational models and memory hierarchies, and creating reliable fine-grained timing mechanisms, as well as others. Successfully demonstrating these channels highlights the possibility of more dangerous side-channel attacks (the presence of a covert channel is a prerequisite for side-channel attacks), as well as providing concrete examples that illuminate the principles for general cross component attacks in heterogeneous systems, expanding our understanding of microarchitectural attacks to guide the development of mitigation strategies in such important systems.

Section~\ref{sec:background} provides an overview of the integrated CPU-GPU systems architecture and our threat model.  We consider two possibilities for creating covert channels in cross component systems: (1) Contention through directly shared microarchitecture resources.  In the case of the integrated CPU-GPU system we use in our experiments, the lowest level of the cache (the LLC) is shared and serves as our example of this type of channel; and (2) Contention through time multiplexed resources such as shared buses, cache ports, computational units and similar resources.  In such resources, if both components use the resource at the same time, there is a perceived delay as their requests contend for the use of the limited resource.  We illustrate this type of channel by building a covert channel attack on the ring-bus interconnect the CPU and GPU to the LLC cache.   We believe another channel type may exist if there are interaction protocols among the components (e.g., for coherence) but we leave exploration of such channels to future work. 

The first attack (described in Section~\ref{sec:covert}) is a PRIME+PROBE based covert channel on the shared Last Level Cache \textit{(LLC)}. Developing this attack requires reverse engineering of the partially documented GPU L3 cache architecture and its interaction with the LLC. Since the GPU cache hierarchy is attached to the CPU hierarchy at the LLC level, we had to work around substantial differences of the cache view including: (1) the index hashing in the GPU L3 is not consistent with the index hashing of the shared LLC: the conflict set addresses to overflow the GPU L3 to cause an LLC access do not hash to the same LLC set, causing a substantial self-interference problem which we mitigated;  (2) the LLC is inclusive on the CPU side, but not on the GPU side limiting some attack strategies; and (3) we had to calibrate the two sides to overcome the disparity of low GPU frequency and high CPU frequency and enable reliable high-quality communication. Another challenge we had to address is the lack of a GPU hardware timer available to time the difference between cache hits and misses:  we developed a custom timer using a kernel that increments a shared variable atomically.  
The second attack we present creates contention among resources due to simultaneous access by CPU and GPU (Section \ref{sec:mem_request}), which also required us to characterize the contention behavior to build reliable and synchronized contention.  The bandwidth and error obtained from the two channels are presented in section \ref{sec:bandwidth}.  
The LLC based channel achieves a bandwidth of 120 kbps with an error rate of 2\% while the contention based channel achieves a bandwidth of 400 kbps with a 0.8\% error rate. We discuss the potential mitigations in Section~\ref{sec:discussion} and compare our work to other related attacks in Section~\ref{sec:related}.

In summary, the contributions of this paper are:
\begin{itemize}
\setlength{\itemsep}{0in}
    \item We present a new class of attacks that span different components within a heterogeneous systems.  
    \item We reverse engineer several components on integrated CPU-GPU system, and develop solutions to challenges relating to cross-component channels.
    \item We illustrate these attacks by building and characterizing two real covert channel attacks on an integrated CPU-GPU system.
\end{itemize}
\noindent
In addition, we believe that these channels provide a valuable first experience with these types of channels which must be considered in the design of secure heterogeneous systems.  Section~\ref{sec:conclusion} discusses possible future research and presents our concluding remarks.


\section{Background and threat model}\label{sec:background}

In this section, we introduce the organization of Intel's integrated GPU systems, to provide background necessary to understand our attack.  We also present the threat model, outlining our assumptions on the attacker's capabilities.


\subsection{Intel Integrated CPU-GPU systems} \label{sec:igpu}
 
 Traditionally, discrete GPUs are connected with the rest of the system through PCIe bus, and have access to a separate physical memory (and therefore memory hierarchy) than that of the CPU. However, starting with Intel's Westemere in 2010, Intel's CPUs have integrated GPUs (\textit{iGPU}, called Ironlake Graphics) incorporated on the same die with the conventional CPU, to support increasingly multi-media heavy workloads without the need for a separate (bulky and power hungry) GPU.  This GPU support has continued to evolve with every generation providing more performance and features; for example the current generation of Intel Graphics (Iris Plus on Gen11 Intel Graphics Technology ~\cite{Intel-Gen11}) offers up to 64 execution units (similar to CUDA cores in Nvidia terminology) and at the highest end, over 1 Teraflops of GPU performance.  Thus, modern processors already use complex System-on-Chip (\textit{SoC}) designs. 
 
 The architectural features and programming interface for the iGPU are similar to those of discrete GPUs in many aspects. For general purpose computing on integrated GPUs, the programmer  uses OpenCL~\cite{opencl} (equivalent to CUDA programming model on Nvidia discrete GPUs~\cite{Nvidia-prog-guide}). Based on the application, programmers launch the required number of threads that are grouped together into work-groups (similar to thread blocks in Nvidia terminology). Work-groups are divided into groups of threads executing Single Instruction Multiple Data (\textit{SIMD}) style in lock step manner (called wavefronts, analogous to warps in Nvidia terminology). 
In integrated GPUs the SIMD width is variable; it changes depending on the register requirements of the kernel.

\cut{The execution model on iGPUs is also similar to that of discrete GPUs. The memory is first allocated for the data to be operated on by the iGPU.  GPU computation expressed as kernels is launched by the CPU and executed by the GPU. Discrete GPUs are equipped with their own fast Graphics Dual Data Rate \textit{(GDDR)} memory that is separate from the system memory. The data is transported to the GDDR from the system memory using DMA, so that the GPU can access and operate on them. As a result, discrete GPUs do not directly share architectural resources with the CPU, making the types of attacks we study not possible. 
On the other hand, }  

\begin{figure}
  \centering
  \includegraphics[width=4cm]{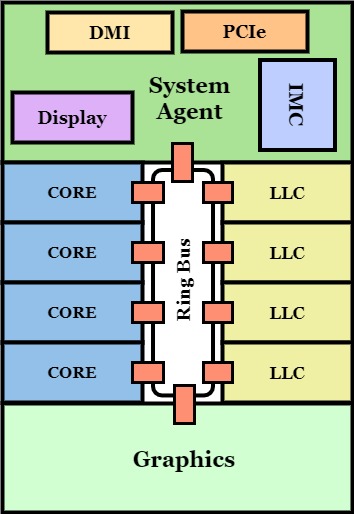}
  \caption{\textbf{Intel SoC architecture}}
  \label{fig:socarch}
\end{figure}

iGPUs reside on the same chip and share the system RAM and connect to the same memory hierarchy as the CPU (typically at the LLC level). Figure~\ref{fig:socarch} shows the architecture of an Intel SoC processor, integrating four CPU cores and an iGPU~\cite{Intel-Gen9}. The iGPU is connected with CPUs and the rest of the system through a ring interconnect: a 32 byte wide bidirectional data bus. The GPU shares the Last Level Cache (\textit{LLC}) with the CPU, which much like the CPU, serves as the last level of the GPUs cache hierarchy. The whole LLC is accessible by the GPU through the ring interconnect with a typical implementation of address ranges hashing to different slices of the LLC. The GPU and CPU can access the LLC simultaneously. However, there is an impact on the access latency if the GPU and CPU contend for accessing, due to factors such as delays in accessing the bus and access limitations on the LLC ports.  We characterize the contention behavior in Section~\ref{sec:mem_request}. The GPU and CPU share other components such as the display controller, the PCIe controller, the optional eDRAM controller and the memory controller.

\begin{figure}
  \centering
  \includegraphics[width=1\linewidth]{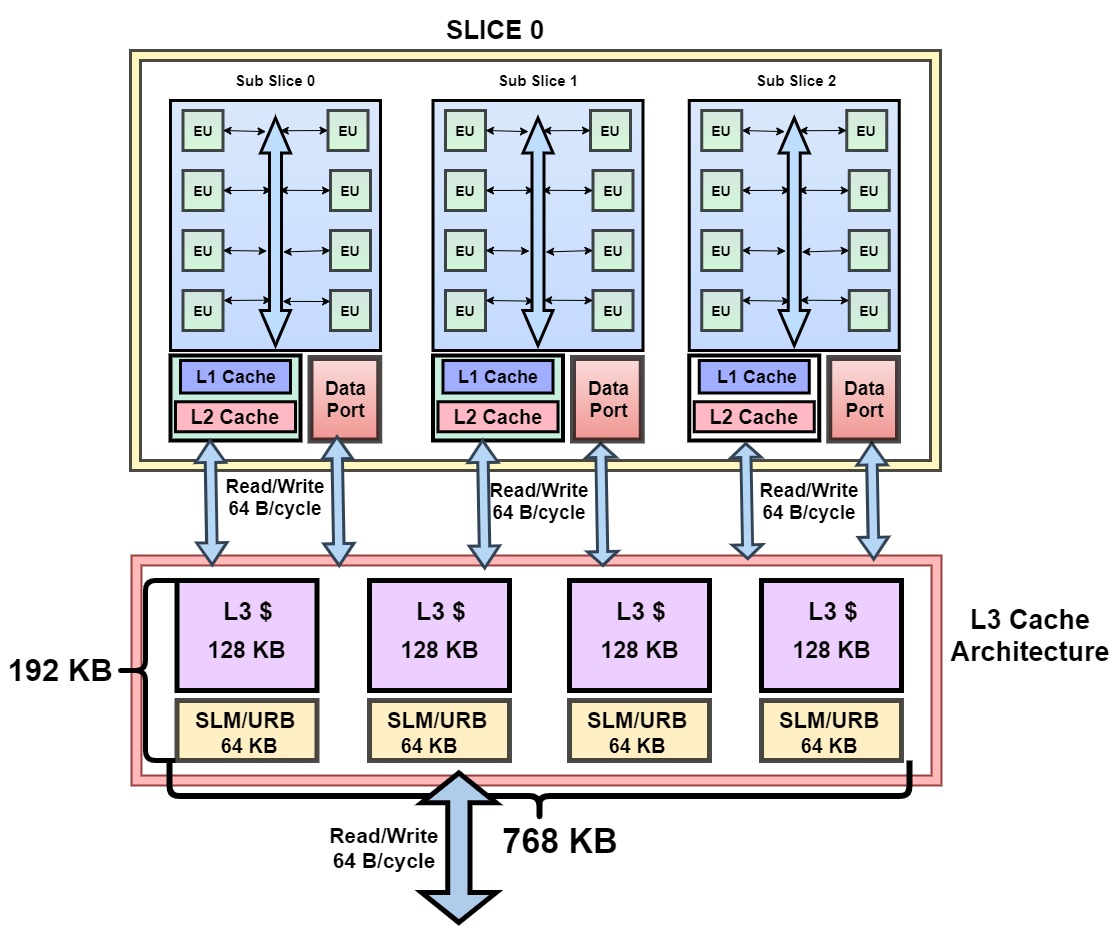}
  \caption{\textbf{Intel integrated GPU architecture}}
  \label{fig:gpuarch}
\end{figure}

The architecture of the iGPU is shown in Figure \ref{fig:gpuarch}. 
A group of 8 EUs (analogous to CUDA cores) are consolidated into a single unit which is called a \textit{Subslice} (similar to SM in Nvidia terminology) and typically 3 subslices create a \textit{Slice}.  The number of slices varies with the particular SoC model even within the same generation, as the slices are designed in a modular fashion allowing different GPU configurations to be created. 
Experimentally, we discovered that multiple work-groups are allocated to different subslices in a round robin manner. The global thread dispatcher launches the work-groups to different subslices. A single SIMD width equivalent number of threads in a single subslice is launched to EUs in a round robin manner as well. 
A fixed functional pipeline in the subslice (not shown in the figure) is dedicated for graphics processing.

The iGPU uses three levels of cache (in addition to the LLC). The first two levels, L1 and L2, are called sampler caches and are used solely for graphics. The third level cache, L3, is universal and can be used for both graphics and computational applications. The L3 cache is common to all the subslices in a single slice. The L3 cache fabric in different slices is also interconnected, giving a consolidated L3 architecture shared by all the EUs in all slices.  We explain the organization of the L3 cache in more detail in Section~\ref{sec:gpu_hierarchy}. In each slice, there is also a shared local memory (SLM), a structure within the L3 complex that supports programmer managed data sharing among threads within the same work-group~\cite{Intel-Gen9}. 


\subsection{Threat Model}
In this paper, we build two covert channel attacks originating from an integrated GPU to the CPU or vice versa. In a covert channel, two distinct processes communicate covertly over a communication channel. The sending process is known as the {\em Trojan}, while the receiving process is called the {\em Spy}. To the best of our knowledge, previously established covert channels were within the same physical device, either a CPU~\cite{maurice2015c5} or GPU~\cite{Hoda-MICRO}, but not spanning both. 
In contrast, our covert channel differs in that the trojan and the spy processes communicate across different heterogeneous components, each featuring a different execution model, memory hierarchy and clock domain. Specifically, the trojan process launches a kernel on the GPU and the spy process operates completely on the CPU during communication. We also demonstrate the communication in the other direction (in fact, we implement bidirectional covert channel). 
We explore two different covert channels, one using a PRIME+PROBE style attack on the LLC, and another that uses contention as the two processes concurrently access a shared resource to implement the communication.

We assume that the trojan and spy processes are both separate user level processes without additional privileges, one running on the GPU and another on CPU. There is no explicit sharing between them (for example sharing of memory objects). The communication on the LLC occurs over pre-agreed sets in the cache. Such agreement is not required in a contention based attack, and can be relaxed by dynamically identifying sets to communicate (but we do not pursue such an implementation). We do not make any assumptions regarding the system environment. All the cores are kept active, with the trojan process running on one CPU (and launching the kernel on the GPU). On the GPU side of our attacks, the program uses the GPU through user-level OpenCL API calls (we suspect that channels could also be established using OpenGL or other graphics calls).  All of our experiments are on a Kaby Lake i7-7700k processor, which features an integrated Intel's Gen9 HD Graphics Neo.  We use OpenCL version 2.0 (Driver version 18.51.12049), running Ubuntu version 16.04 LTS (which uses Linux Kernel version 4.13).  The attacks were developed and tested on an unmodified but generally quiet system (not running additional workloads) on the GPU side of the attack. Current iGPUs are not capable of running multiple computation kernels from separate contexts concurrently and therefore no noise is expected on the GPU side. If future generations of iGPUs allow sharing, then some of the strategies used in discrete GPU attacks~\cite{Hoda-MICRO} could be leveraged.  
On the CPU side of the attack, no additional constraints were made. We introduce techniques to tolerate noise by utilizing multiple redundant cache sets, and by tuning the overlap between the GPU and CPU to increase the signal. 
\section{LLC-based Covert Channel}\label{sec:covert}

This section presents the first covert channel attack: a Prime+Probe channel using the shared LLC cache.  Prime+Probe is one of the most common strategies of cache-based attacks~\cite{percival-05}; it is also one of the most general strategies because it does not require sharing of parts of the address space as required by other strategies, for example those requiring sharing to be able to flush data out of the caches.   In Prime+Probe, first the spy process accesses its own data and fills up the cache (priming).  Next, the trojan either accesses its own data (replacing the Spy's) to communicate a "1", or does nothing to communicate a "0".   Finally, the spy can detect this transferred bit by re-accessing its data (probe) and measuring the access time.  If the time is high, indicating a cache miss, it detects a "1", otherwise a "0".   


\subsection{Attack Overview and Challenges}\label{sec:cache_attack_ov}

\begin{figure}[!htbp]
  \includegraphics[width=\linewidth]{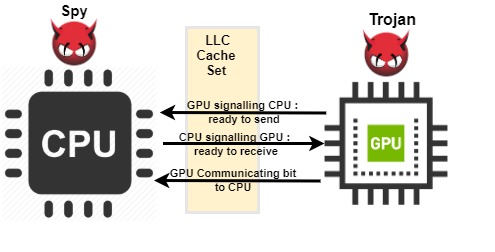}
  \caption{\textbf{LLC based CPU-GPU covert channel overview }}
  \label{fig:cov_chan_ov}
  \vspace{-2mm}
\end{figure}

In this attack, the CPU and GPU communicate over the LLC cache sets. Figure \ref{fig:cov_chan_ov} depicts the overview of the attack. We illustrate the attack at a high level using a trojan process launched on the GPU, communicating the bits to the CPU but the opposite is also possible. The Spy process which is receiving the bits is launched on the CPU. Communication from GPU to CPU is a 3 step process. The first two steps are for handshaking before the communication to make sure that the two sides are synchronized, which is especially important for heterogeneous components that can have highly disparate communication rates. The GPU initiates the handshake by priming the pre-agreed cache set and letting the CPU know that it is ready to send.  Once the CPU receives the signal by probing the same cache set, the CPU acknowledges it back by priming a different cache set and sending ready to receive signal back to GPU in the second phase. GPU receives ready to receive signal by probing the same cache set that was primed by CPU. This ends the handshaking phase and the attack moves to the third step, when GPU sends the data bit to CPU. For sending 1, GPU primes the LLC cache set that is probed by CPU.  If GPU wants to send 0, it doesn't prime the cache set but CPU still probes. This 3 phase communication repeats communicating the secret bits covertly from the GPU process to the CPU process.  

\label{sec:llc_challenges}

Although at a high level this attack strategy is similar to other covert channel attacks, there are a number of unique challenges that occur when we try to implement the channel between the CPU and GPU.  The challenges generally arise from the heterogeneous nature of the computational models on the two components, as well as the different memory hierarchies they have before the shared LLC.  We overview these challenges next.  

\begin{itemize}
\setlength{\itemsep}{0in}
   \item {\bf Absence of a GPU timer:} Prime+Probe attacks rely on the ability to time the difference between a cache hit and a cache miss to implement communication.  
   Usually, a user level hardware counter is available on the system to measure the access latency.  While this is true on the CPU side, unfortunately OpenCL on iGPUs does not provide any such means to the programmer. We describe this problem and the custom user level timer we develop to overcome it in Section~\ref{sec:timer}. 

     \item {\bf Reverse Engineering the LLC viewed from the GPU}
To be able to target specific sets in the LLC for covert communication, we require the knowledge of the physical addresses mapping to cache addresses from both CPU and GPU side (the LLC is physically indexed). Modern GPUs come with their own page tables and paging mechanisms. In Section~\ref{sec:LLC_ev}, we describe how we use the mechanism of shared virtual memory ~\cite{SVM} and zero copy memory to maintain the same physical and virtual addresses across the device.   When a CPU process initializes and launches the GPU kernel, the CPU page table is shared with the GPU in this scenario. This sharing allows us to reverse engineer the cache from the CPU using established techniques~\cite{yarom2015} and use these results on the GPU. 

   \item {\bf Reverse engineering the GPU cache hierarchy:}  While the Intel CPU cache hierarchy is well understood, the GPU cache hierarchy details are not published.  It is critical to understand the cache hierarchy since it determines how memory accesses spill over to the LLC where the covert channel is being implemented.   Since L1 and L2 caches are not used by OpenCL, we need to reverse engineer the GPU L3 to understand how to control the memory references that are evicted from it.  First, we needed to understand whether the LLC is inclusive of the L3 which would make simplify eviction from the L3 from the CPU side.  However, we discover that it is not inclusive, which requires us to understand the L3 in detail in order to control evictions from it.  We describe this challenge in Section~\ref{sec:gpu_hierarchy}.

\end{itemize}

\subsection{Building Custom Timer}\label{sec:timer}

Access to a high-resolution timer is essential to the ability to carry out cache based covert channels; without it we are unable to discriminate a cache hit from a cache miss, which is the primary phenomena used in the communication. 
Although Intel based integrated GPUs have a timer, by default, the manufacturer does not provide an interface to query it in OpenCL based applications. OpenCL programs executing on Intel devices are compiled using the Intel graphics compiler \textit{(IGC)} ~\cite{IGC-19,IGC-git}. In debug mode, it is possible to query an overloaded timer function in the program. This is not available to the programmer in default mode and requires a superuser permission for installation. In our end-to-end covert channel threat model, the attacker has no privileged access. 
Therefore, we need to come up with an alternative approach to measure the access latency within the GPU application. 

We leverage GPU parallelism and hardware Shared Local Memory \textit{(SLM)} to build the custom timer. Shared local memory in Intel based iGPUs is a memory structure, shared across all EUs in a subslice. 64 Kbytes of shared local memory is available per subslice. Shared local memory is private to all the threads from a single work-group. We launch a work-group for which certain number of threads are used to conduct the attack and the rest of the threads are used to increment a counter value stored in shared memory. The threads that are responsible for carrying out the attack read the shared value as timestamps before and after the access to measure the access time (the principle of this technique was used in CPU attacks on the ARM where the hardware time is not available in user mode~\cite{ARMMagedon}). Due to branch divergence within the wavefronts (SIMD width of threads), the execution of two groups of threads in a single wavefront gets serialized. So the number of threads that are used for counter increment start at a wavefront boundary till the end of the work-group. 
Each LLC cache set consists of 16 ways which can be probed in parallel from the GPU using 16 threads (thread id 0 - 15). But the timer should start from wavefront boundary \textit{i.e.} above 32 threads (thread ID$>$31) (In our case, the wavefront size is 32). So the threads involved in conducting the attack is from 0 to 15 and the threads involved in the counter increment is from 32 and above, till the end of work-group. Ideally only 2 wavefronts can be used where the first wavefront is responsible for attack and the second wavefront used for counter. However, we found out that the timer resolution obtained by using a single wavefront is not adequate to distinguish between access latency of different memory hierarchy levels. To obtain a desired timer accuracy we launched the maximum number of threads allowed in a work-group (256) of which 224 threads were used for the counter, and the remaining 32 for memory accesses.


\begin{small}
\begin{algorithm}
\SetAlgoLined
\textit{volatile} \textunderscore\textunderscore\textit{local} counter\\ \label{timerLn1}
\textit{cl\textunderscore uint} start,end,idxVal\\
\textit{cl\textunderscore ulong} average\\
\textit{cl \textunderscore float} access\textunderscore time\\
\uIf{thread\textunderscore ID$>$SIMD length}{\label{timerInc1}
    \For{$i = 0;\ i < n;\ i = i + 1$}{ 
    \textit{atomic\textunderscore add}(counter,$1$)\\
  }\label{timerInc2}
}
\uElse{ \label{timerRead1}
    \tcc{Measure time over \textit{x} accesses }
     average$ = 0$\\    \label{timerLn2}
    \For{$i = 0;\ i < x;\ i = i + 1$}{
        
       start = \textit{atomic\textunderscore add}(counter,0)\\
       idxVal = data\textunderscore buffer[idxVal]\\
       end = \textit{atomic\textunderscore add}(counter,0)\\
       average $+=$ end $-$ start\\
    }\label{timerRead2}
    access\textunderscore time = (\textit{cl \textunderscore float})(average/\textit{x})\\   \label{timerLn3}
    \tcc{Clear data from L3 but not LLC}
    \tcc{Repeat  \ref{timerLn2} to \ref{timerLn3} for LLC access} \label{timerLn4}
    \tcc{Repeat \ref{timerLn2} to \ref{timerLn3} again for L3 access}\label{timerLn5}
     
}

\caption{Custom Timer Algorithm}

\label{alg:algoTimer}

\end{algorithm}

\end{small}

\begin{figure}[ht]
\centering
\vspace{-3mm}
\includegraphics[width=0.9\linewidth]{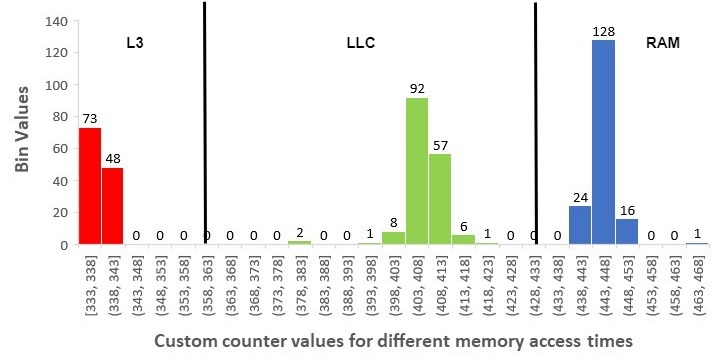}
  \caption{\textbf{Custom Timer Characterization}}
  \label{fig:cust_timer}
    \vspace{-4mm}
\end{figure}

Algorithm \ref{alg:algoTimer} demonstrates the custom timer code inside the GPU kernel.  Data accessed from the iGPU, using OpenCL, can get cached into the LLC and the L3. To conduct a covert channel attack, the attacker needs to distinguish 3 levels of access time, \textit{i.e.} system memory, LLC and L3. In line \ref{timerLn1} of algorithm \ref{alg:algoTimer}, variable \textit{volatile \textunderscore\textunderscore local counter} is declared, which is used as the timer. The \textit{volatile} keyword makes sure that the counter variable is not cached inside the thread's registers. The timer variable is declared in the shared memory of the device using \textunderscore\textunderscore\textit{local} keyword. Shared memory uses a separate data path than that used for accessing L3, which makes sure that there is no resource contention that can lead to erratic counter updates. To test the custom timer, we launched a kernel with 1 work-group consisting of \textit{max} number of threads per work-group. Threads over a single wavefront are used to increment the counter atomically as shown in lines \ref{timerInc1} - \ref{timerInc2} in the \textit{if} section of the algorithm. Atomic operation on the variable ensures that the variable is accessed and incremented properly. In lines \ref{timerRead1} - \ref{timerRead2}, the data is accessed and the value of the counter is read atomically. A number (\textit{x}) of memory accesses is timed and averaged. The first access represents the measurement from the system memory. To measure the access time from the LLC, the data is cleared from the L3 but made sure that it is not cleared from the LLC and then \ref{timerInc1} - \ref{timerLn3} is repeated to measure the access time from LLC. Now as the data is both cached in LLC and L3, repeating steps \ref{timerInc1} - \ref{timerLn3} yields the L3 access time. 

Figure \ref{fig:cust_timer} shows our experiment with the timer measuring access time from the different levels of the hierarchy (shown in different colors). The access times obtained from the counter values are clearly separated enabling us to distinguish between accesses from the three levels of hierarchy.  

\subsection{Reverse engineering the LLC}\label{sec:LLC_ev}

 The next challenge in the attack is the the formation of eviction sets which are used in both the prime and probe steps to occupy the cache sets targeted for communication.  
The eviction set is a set of physical addresses that mapped onto the same cache set ~\cite{vila2019theory}. Once the attacker acquires the addresses that are in the same cache set, she can monitor the victim's activity by manipulating the cache set. 
We discuss this challenge first from the CPU side, leveraging techniques developed from previous attacks to identify the conflict set.  Then, we discuss how to create this set for the GPU side, by leveraging OpenCL Shared Virtual Memory feature. 

\noindent
{\bf Deriving LLC conflict sets from the CPU:} 
To procure the addresses mapped to the same set, the attacker must reverse engineer (or otherwise know) the cache configuration.  She also needs to know about the virtual to physical address mapping since the LLC is physically addressed. 
Modern LLCs are divided into a number of slices that varies with the processor architecture. The cache slice selection depends on a complex index hashing scheme designed to evenly distribute the addresses across the slices. 
The Intel architecture that we are using has 8 MB last level cache divided into 4 slices of 2 MB each. The cache is 16-way set associative, with 64-byte cache lines (a total of 2048 cache sets per slice). 
Slice selection uses a complex hash function that is not revealed by the vendor. Previously conducted attacks~\cite{irazoqui2015,maurice2015,yarom2015,mehmet-2016} in the LLC have reverse engineered this complex addressing scheme.   We used a similar approach to reverse engineer the index hashing including the use of huge memory pages (1GB).  On our processor, we discover that the index hashing algorithm selects a slice using 2 bits computed as follows.

\begin{equation}
\begin{aligned}\label{eq:0_1}
S_{0} = b_{36}\oplus b_{35} \oplus b_{33} \oplus b_{32} \oplus b_{30} \oplus b_{28} \oplus b_{27} \oplus b_{26}\\ \oplus b_{25} \oplus b_{24} \oplus b_{22} \oplus b_{20} \oplus b_{18} \oplus b_{17} \oplus b_{16}\\ \oplus b_{14} \oplus b_{12} \oplus b_{10} \oplus b_{6}\\
\end{aligned}
\end{equation}

\begin{equation}
\begin{aligned}\label{eq:0_2}
 S_{1} = b_{37}\oplus b_{35} \oplus b_{34} \oplus b_{33} \oplus b_{31} \oplus b_{29} \oplus b_{28} \oplus b_{26}\\ \oplus b_{24} \oplus b_{23} \oplus b_{22} \oplus b_{21} \oplus b_{20} \oplus b_{19} \oplus b_{17}\\ \oplus b_{15} \oplus b_{13} \oplus b_{11} \oplus b_{7}\\
\end{aligned}
\end{equation}

\noindent 
{\bf GPU LLC Conflict Sets:} The next challenge is how to derive a conflict set from the GPU side: the GPU has it's own paging mechanism ~\cite{Intel-GPU-PRM-Vol-5} that is different from the CPU paging mechanism. Therefore, we need to form the LLC eviction set from GPU side as well. However, GPU computing using OpenCL on intel GPUs allows the programmer to allocate memory with the same virtual address space using \textit{Shared Virtual Memory} (SVM)~\cite{SVM} and the same physical address space through zero copy buffers~\cite{Intel-zero} from the user level by using APIs provided in OpenCL. So when a CPU process initializes and launches the GPU kernel, on shared pages the eviction set identified from the CPU side also holds for the GPU after the GPU kernel is launched.   Please note that this sharing is within the address space of the process launching the GPU side of the attack; no sharing is required between the Spy and Trojan.  

\subsection{Reverse engineering GPU Caches}\label{sec:gpu_hierarchy}

Reverse engineering the LLC allows us to understand what access patterns presented to the LLC is needed to carry out the attack.  However, the GPU cannot simply generate those addresses since they could be cached in the GPU cache hierarchy and as a result never spill over to access the LLC.   On the Intel iGPUs, the L1 and L2 caches are used only for graphics workloads.   Thus, to be able to create the reference pattern from the GPU that will result in the eviction set address patterns to access the LLC, we must first reverse engineer the L3 cache on the GPU: if we understand the organization of the L3, we can design a memory reference pattern that causes the desired LLC accesses to occur.  

Most of previous attacks on last level caches depend on the cache inclusiveness property~\cite{yarom2015}. With inclusive caches, data evicted from the lower level of caches also gets evicted from the higher level caches. 
As a first step of understanding the L3, we first determine whether it is indeed inclusive (we discover that it is not).  Next, we reverse engineer the structure of the cache, and finally, we develop conflict sets that allow us to control the traffic that gets presented to the LLC.  


\noindent
{\bf L3 inclusiveness:} 
To check whether the L3 is inclusive, we design the following experiment. 
We create a buffer shared by the CPU and the GPU. We identify a set of \textit{n} addresses which are accessed first by the GPU.  Initially,  the caches are cold, and the data is brought from memory and cached in both LLC and L3. Next, the CPU accesses the same data bringing it into its caches and then flushes the data removing it from all the cache levels using \textit{clflush}. If the LLC is inclusive of the L3 cache, the removal of the flushed data from the LLC will cause back-invalidations to evict the data from the L3 cache of the GPU as well. Finally, we check whether the data is still present in the L3, by accessing the data from the GPU side and timing it using our user timer.  Based on access time, we observed that the data is accessed from L3 and not from the memory, indicating that the L3 cache is not inclusive. 



\noindent
{\bf L3 Architecture Details ~\cite{Intel-GPU-PRM-Vol-7}:} 
Figure~\ref{fig:gpuarch} earlier in the paper shows the L3 within the iGPU hierarchy. \cut{Figure \ref{fig:L3Arch} shows the details of the L3 internal architecture.} The total L3 cache capacity may vary from one GPU generation to another. Irrespective of the total cache size, each slice of the iGPU is accompanied by its own L3 cache slice of size 768 KB. Each L3 cache slice is further divided into 4 cache banks, each consisting of 192 KB. This 192 KB is configured to 128 KB of L3 cache and 64 KB of Shared Local Memory \textit{(SLM)}. 

The critical path to access the SLM is separate from the L3 access path. Accessing the SLM does not impact adversely L3 access latency and vice versa, a feature that makes it possible to implement our user level timer measurement without interference from the memory traffic.
The L3 uses a tree based pLRU based cache replacement policy with $(N-1)$ number of nodes in the tree, N being the number of ways in the cache set ~\cite{Intel-GPU-PRM-Vol-7}. 
\cut{
\begin{figure}[!htbp]
  \includegraphics[width=\linewidth, height=0.6\linewidth]{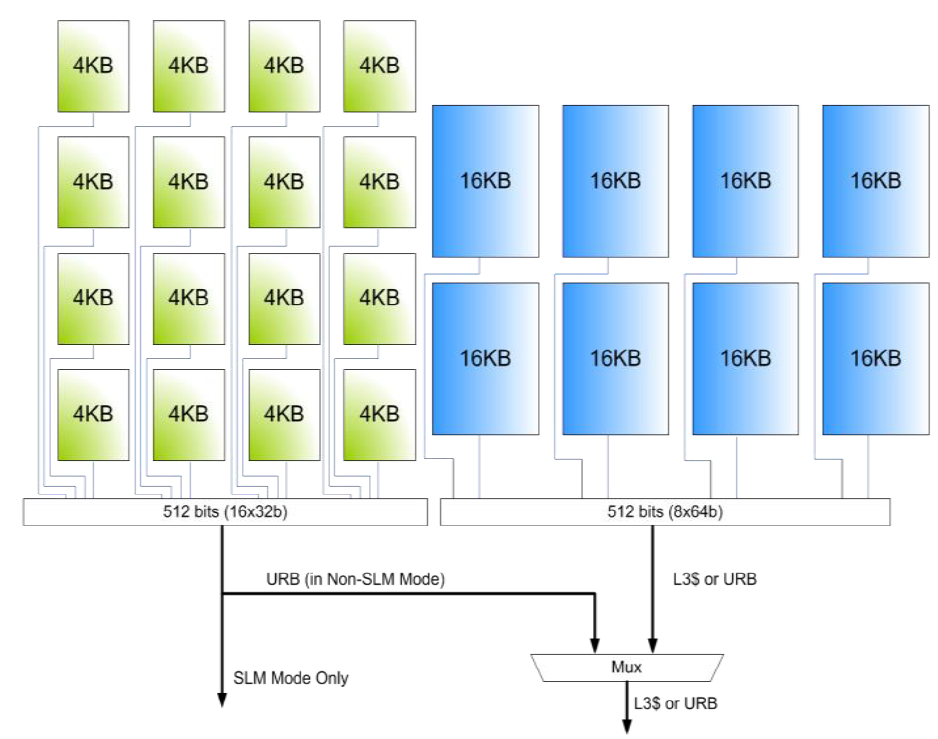}
  \caption{\textbf{GPU L3 architecture}~\protect\cite{Intel-GPU-PRM-Vol-7}}
  \label{fig:L3Arch}
\end{figure}
}

\noindent
{\bf L3 Eviction Set:}
Although~\cite{Intel-GPU-PRM-Vol-7} provides some architectural details of the L3 structure, architectural information critical to carry out a timing based PRIME+PROBE attack is still missing. Given that the L3 is non-inclusive, we cannot evict it from the CPU side, and instead need to create an eviction set from the GPU side to conduct a successful attack. 
In order to form an eviction set in a particular cache level, an attacker requires cache details like the cache line size, number of cache lines in a set and number of cache sets. Also understanding the mapping of an address to a cache set is required to acquire the addresses that are mapped to the same cache set. After figuring out the addresses placed in the same cache set, an attacker also considers the cache replacement policy to evict the target address from the cache set. The total \textit{Cache Size} is the product of the cache line size (\textit{CL$_{S}$}), number of cache lines per cache set (\textit{N$_{L}$}) and the number of cache sets (\textit{N$_{S}$}).


The understanding of L3 cache mapping is required to figure out the eviction set. We reverse engineer the configuration of the L3 and discover that the cache line size is 64B.  GPUs are byte addressable and 6 bits in the address represent the byte offset in the cache line. We identify that there are 64 cache lines per cache set, with each cache set spanning 4 KBs. However, the L3 cache is partitioned into 4 banks and each bank is again partitioned into 8 sub-banks. The number of sets per cache bank is 32 which requires 5 bits in the address bits for mapping. There are 4 cache banks which require additional 2 bits in the address for mapping. Each cache bank is again divided into 8 cache sub-banks which require additional 3 bits in the address. So the total of 16 bits (6 bits byte offset + 5 bits for cache set + 2 bits for cache bank + 3 bits for sub-banks) in the address are required for the placement of the data in the L3 cache. We assumed a low order address interleaving which defines the 16 LSB bits. To verify the eviction set, we gathered the addresses with same 16 bits in the LSB and conducted the eviction set test. As the replacement policy is pseudo-LRU (pLRU), accessing the other addresses multiple times (5 times or more in our experiments) guarantees stable eviction of the target address through the pLRU.

During the attack phase on the LLC, we start with the addresses that are in the same LLC set (selected from the LLC eviction set). In both, the priming and the probing phases, these addresses need to be evicted from the L3 so that they are received by the LLC to implement the prime and probe at that level. To evict these target addresses from the L3, we create an eviction set from addresses that have the same last 16 bits as the target address, which in turn ensures that the target and the eviction addresses are in the same L3 cache set. We are careful to choose eviction addresses for the L3 from sets that are not our target at the LLC level.  Otherwise, if the evict and target addresses lie in the same set at the LLC level, these evicted addresses add noise to the LLC set interfering with the covert communication.


\subsection{Putting it all together--LLC Channel}\label{sec:cache_attack_det}

\begin{figure}[!htbp]
 \includegraphics[width=\linewidth]{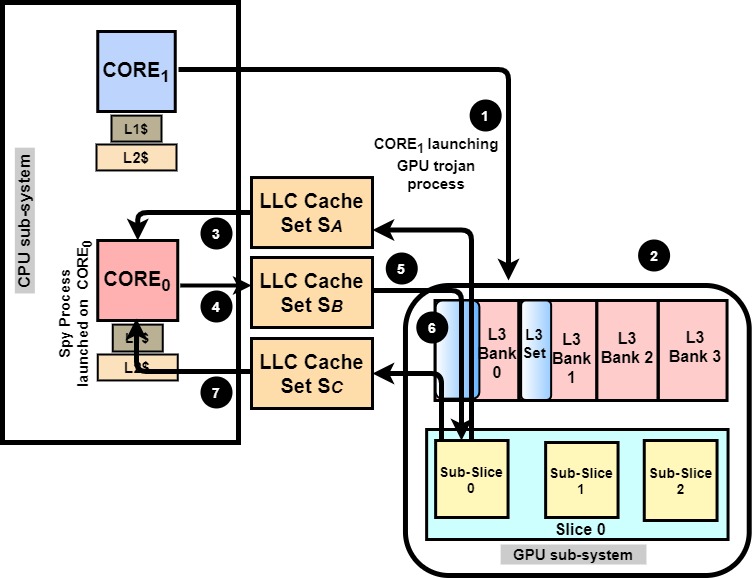}
  \caption{\textbf{LLC based CPU-GPU covert channel details }}
  \label{fig:cov_chan_det}
 \vspace{-3mm}
\end{figure}

By addressing all the challenges, we are to proceed with the covert channel creation on LLC. In our attack model, the spy process is launched on the CPU side (CORE 0). As shown in Figure~\ref{fig:cov_chan_det}, CPU CORE 1 launches the GPU trojan process in step \circled{1} of the attack. On the GPU side of the attack, the eviction set, both on L3 and LLC level is determined before launching the attack.  Before each bit transfer, a handshaking takes place in steps \circled{2} - \circled{5} to ensure synchronization of spy and trojan. The actual bit transmission is done in steps \circled{6} and \circled{7}. A separate LLC cache set is used in each phase of the attack. LLC set S$_{A}$ and S$_{B}$ are used for the handshaking and set S$_{C}$ is used for communication. To conduct the attack, we launched one work-group that is allocated to a sub-slice. The implementation requires synchronization between threads which can only be obtained within a work-group.  In principle, this process can be replicated over multiple work-groups to scale bandwidth.

The GPU initiates the handshaking as data is transferred from GPU to CPU. In the first phase of the handshaking, the GPU trojan process informs the CPU spy process that it is ready to send the data. Step \circled{2} indicates that GPU primes LLC set S$_{A}$ and then probing is done from the CPU side as shown in step \circled{3}. To ensure the GPU accesses the data from the LLC and not the L3, those target addresses need to be evicted from the L3 due to the non-inclusive property of the L3, as shown in Section~\ref{sec:gpu_hierarchy}. Eviction set creation on L3 level is required to evict the target addresses from the L3. We used a separate pollute buffer to evict the target addresses from L3. The pollute buffer addresses and the target addresses are in the same L3 cache set, but lie in separate LLC sets to avoid polluting the sets being used for the attack. After GPU priming is over, the CPU \textit{spy} process probes the same set S$_{A}$ as shown in \circled{3}. The higher level CPU L1 and L2 caches are inclusive of the LLC. So after the GPU \textit{trojan} process finishes priming, the data gets evicted from all levels of caches on the CPU \textit{spy} process. As a result, the subsequent accesses reaches the LLC from the CPU.


The second phase of the handshaking indicates to the GPU \textit{trojan} process by the CPU \textit{spy} process that it is ready to receive. In this phase, the CPU primes the LLC set S$_{B}$ in step \circled{4}. Priming from the CPU on LLC requires creation of the eviction set on the LLC level. The GPU then probes the same cache set S$_{B}$ in step \circled{5}. Probing on the LLC from the GPU side requires eviction on the L3 level again. So another eviction set is determined for a successful second handshaking phase. In this phase of handshaking, the access delay is measured from the GPU side. 
We use our custom timer to measure the delay as described in subsection ~\ref{sec:timer}.

After the whole handshaking phase is over, the actual bit transmission occurs over LLC set S$_{C}$ as shown in steps \circled{6} and \circled{7}. The priming step \circled{6} on the GPU side is similar to step \circled{2} in the first phase of the handshaking. The probing step \circled{7} on the CPU side is similar to step \circled{3}. Similar to previous steps , special consideration for L3 and LLC handling is required. The actual bit transmission takes place if the handshaking phase is successful.
Step \circled{1} is conducted once to launch the kernel on the GPU side. Steps \circled{2} - \circled{7} are conducted within the kernel in a  \textit{for all} loop for the number of bits that are required to be transferred.  

We also built a reverse channel where the Trojan is on the CPU communicating to a Spy on the GPU.  The technical details and overflow of the attack are similar to the opposite direction channel described above, but with the roles reversed.  Specifically, the CPU initiates the handshake by priming set S$_{A}$ while the GPU receives it by probing the same set. Next, the GPU sends a ready to receive signal by priming set S$_{B}$ and the CPU probes the same set to receive it. Finally, the CPU sends the communication bit to the GPU using set S$_{C}$.

\cut{Since the spy and the trojan use completely different computation models operating at substantially different clock rates, determining how to best implement the channel to improve bandwidth and reduce noise is tricky.} 



\noindent 
{\bf Optimization around heterogeneous components:}
The CPU and GPU operate with different clock domains. The iGPU uses a clock rate of 1.1 GHz that is one-fourth of the CPU clock rate 4.2 Ghz (not considering DVFS, which we did not observe on our desktop machine). This  frequency  imbalance  leads to loss of coherence and inability to communicate. By the time GPU primes an LLC set, CPU could already be finished probing leading to a missed communication.   This affected both the bandwidth and reliability of our initial implementation.
To overcome this complication, we take the advantage of GPU thread level parallelism. We observed that as we increase the number of GPU threads, it reduces the frequency disparity between the prime and probe rates on the two sides. While the CPU primes/probes the LLC cache lines in a set serially, the slower GPU can match the cache access rate by operating in parallel.  

\section{Contention Covert Channel}\label{sec:mem_request}

Even with absent direct sharing of stateful microarchitectural components (such as the LLC), contention may arise when the two components share a bandwidth or capacity limited microarchitectural structure such as buses or ports.  In such situations, measurable contention can also be achieved if the two processes running on the two components access the same structure concurrently (observing slow downs).  Although there are likely to be a number of such shared contention domains on our system, we implement an attack based on contention on the ring bus connecting the CPU and GPU to the LLC.  
Specifically, when both the CPU and GPU generate traffic to the LLC, they each observe delays higher than when only one of them does, providing a way to communicate two states by either creating contention or not.  

Since contention relates to concurrent use of the shared resource, it requires accurate synchronization between the two sides, which is challenging in the presence of the clock frequency disparity between CPU and GPU. The CPU runs at 4x the speed of the GPU and the data access delay cannot be observed if the GPU data access is lower than a limit. Through our systematic study, we identified the parameters that contribute in creating a robust contention based channel with low error rate and high bandwidth. We also devised a parameter that controls the frequency disparity between the computational resources.  We describe the attack in more detail in the remainder of this section.

\noindent\textbf{Attack Overview:}
The attack creates contention on the ring bus between the CPU and GPU used to access the LLC.   During the attack, the CPU and GPU generate addresses chosen from their own pre-allocated memory buffers.  The CPU and GPU buffers are chosen to map to different  LLC sets to avoid LLC conflicts distorting the contention signal. 
With the two processes accessing disjoint sets in the cache, the contention occurs strictly on the shared resources leading to the LLC. 

\begin{figure}[!htbp]
\vspace{-3mm}
\centering
  \includegraphics[width=0.8\linewidth]{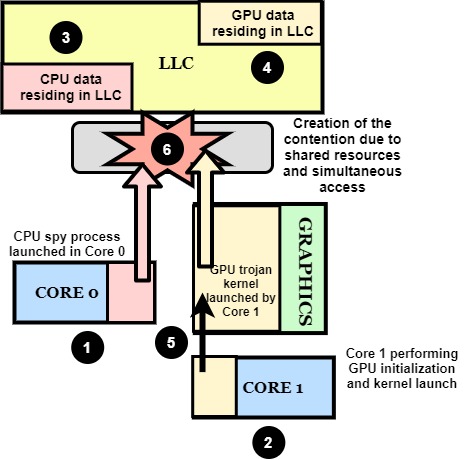}
  \caption{\textbf{Contention channel attack methodology}}
  \label{fig:busCov_1}
  \vspace{-3mm}
\end{figure}

The attack overview is present in Figure \ref{fig:busCov_1}.  The CPU process is launched in CORE 0 and a GPU process is launched in CORE 1 as shown in step \circled{1} and \circled{2}. 
The processes launch each carries out data allocation and initialization. The trojan process launched on CORE 1 launches the GPU kernel as shown in step \circled{3}. The data is accessed by the CPU and GPU simultaneously. The first access will warm-up the cache and bring the CPU and GPU data to the LLC, steps \circled{4} and \circled{5}. Subsequent memory accesses would hit the LLC and generate contention among the shared resources as shown in step \circled{6}. This contention among the shared resources gets reflected during the data access by the CPU. 

\noindent\textbf{Contention Channel Implementation:}
To build the covert channel, we need to identify different parameters that contribute towards building the channel to be able to systematically create and optimize the attack.    For the CPU, \textit{T$_{CPU}$} is the time required to access \textit{S$_{CPU}$} bytes of data. With the simultaneous access from the GPU, the access time is increased by \textit{T$_{OV}$}. The total time \textit{T$_{TOTAL_{CPU}}$} required to access the data from the CPU during the simultaneous GPU access is given in Equation \ref{eq:1_1}. The overhead created due to simultaneous access is a function of the \textit{S$_{GPU}$} bytes of data accessed by GPU, number of threads launched \textit{NUM$_{Threads}$} and an Iteration Factor factor $I_{F}$ reflecting how many iterations the data is accessed as shown in Equation \ref{eq:1_2}. One constraint is to keep both, CPU and GPU data, in the last level cache; the the total of \textit{S$_{GPU}$} and \textit{S$_{CPU}$} has to be less than the total size of the last level cache, as shown in Equation \ref{eq:1_3}. Another constraint is that the LLC sets that are mapped to the CPU buffer should not coincide with the sets that are mapped to the GPU buffer, as shown in equation \ref{eq:1_4}.
\begin{gather}
T_{TOTAL_{CPU}} = T_{CPU} + T_{OV}\label{eq:1_1}\\
T_{OV} = f(I_{F}).f(S_{GPU}).f(NUM_{Threads})\label{eq:1_2}\\
\textrm{s.t.} \hspace{2ex} S_{CPU} + S_{GPU}\ll S_{LLC}\label{eq:1_3}\\
S_{CPU} \cap S_{GPU} = \emptyset\label{eq:1_4}
\end{gather}
On the CPU side of the attack, a buffer size of \textit{S$_{CPU}$} bytes has been created. The accesses are done at an offset of cache line size of 64b. So the number of accesses are equivalent to the number of cache lines in the allocated buffer. The data is accessed in a random pointer chasing manner to lower prefetching effects that may cause replacements of either the CPU or GPU data in the LLC. 
First, LLC is warmed up. The subsequent accesses would be serviced from the LLC.
The size of the buffer is chosen to ensure that the data is evicted from local caches but not from the LLC. Each access time is measured by \textit{clock\_gettime()}.

On the GPU side, the number of cache lines needed to be accessed is divided among the number of threads launched. The number of memory addresses that each thread needs to access, \textit{numElsPerThread}, is shown in in Equation \ref{eq:2_1}. 

 \begin{equation}
 \begin{aligned}\label{eq:2_1}
\textit{numElsPerThread} = \dfrac{\textit{number of cache lines}}{\textit{number of threads}}
 \end{aligned}
 \end{equation}

One of the novel problems presented by asymmetric covert channels is that the two sides have an asymmetric view of the resource; for example, the GPU and CPU operate at different frequencies, and the GPU must overflow the L3 cache to generate an access to the LLC, which unlike the CPU side requires deriving different conflict sets due to the different indexing scheme.  Without calibration, this mismatch can lead to inefficient communication, reducing bandwidth and increasing errors. We introduce the notion \textit{Iteration Factor} $I_{F}$ to allow us to align the two ends of the channel as shown in equation \ref{eq:1_2}. For a given GPU buffer size, the execution time varies based on the number of work-groups launched. $I_F$ (the number of iterations the data is accessed on the GPU) ensures that the ratio between the GPU and CPU execution time is near 1. 

\section{Evaluation}\label{sec:bandwidth}
In this section, we evaluate the two covert channels in terms of channel bandwidth and error rate.

\begin{figure}[ht]
\vspace{-3mm}
\centering
  \includegraphics[width=\linewidth]{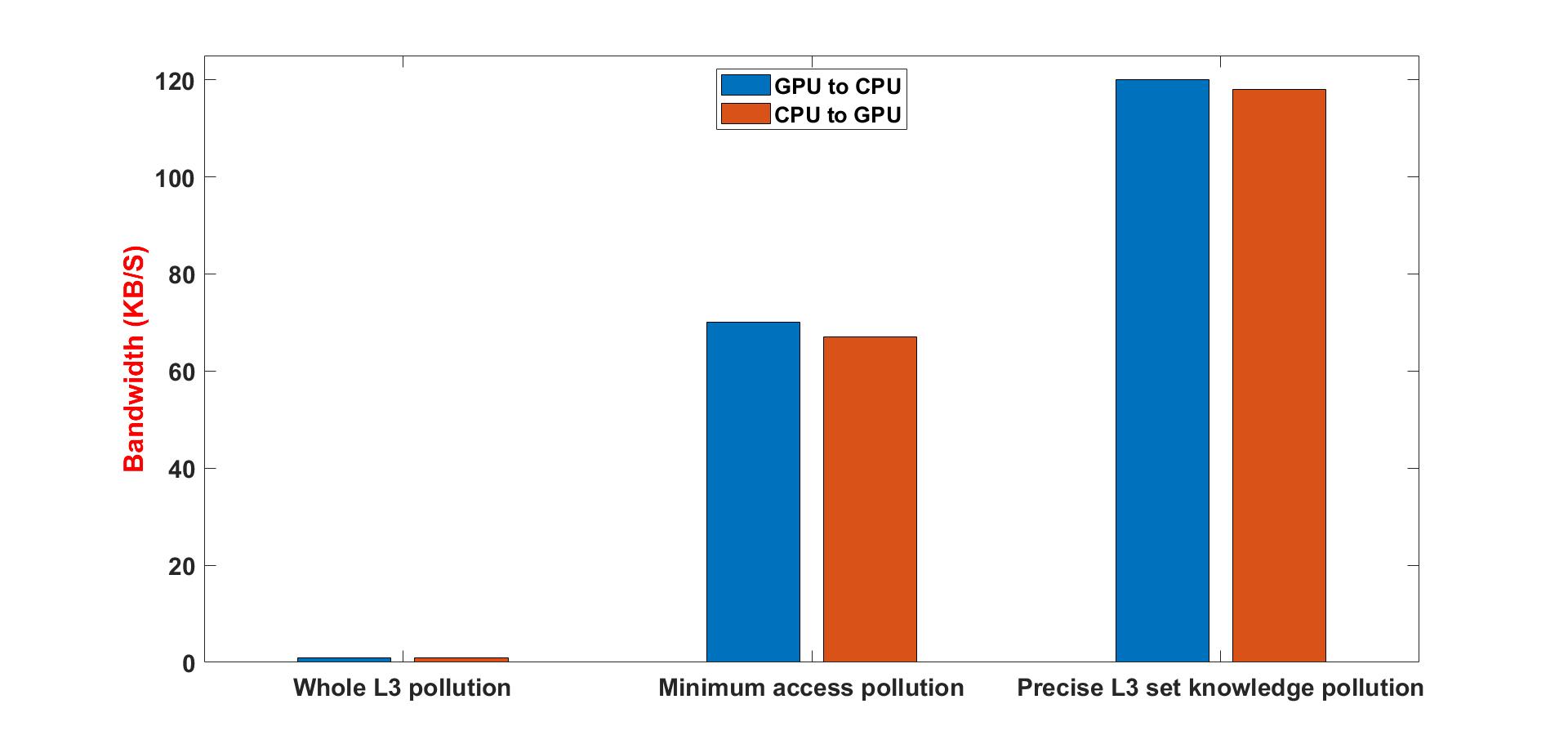}
\caption{\textbf{LLC bandwidth (different L3 eviction strategies)}}
\vspace{-0.15in}
\label{fig:LLC_Cov_1}
\end{figure}

\noindent \textbf{LLC-based Covert Channel:}
The GPU L3 cache is non-inclusive which requires it to be filled to overflow and access the LLC.  Figure \ref{fig:LLC_Cov_1} shows the bandwidth of the channel on both directions (CPU-to-GPU and GPU-to-CPU channels) based on different strategies to overflow the L3. The naive way to establish the covert channel can be performed by clearing the whole L3 cache (we can use the GPU parallelism to accelerate this process); the advantage here is that we do not have to reverse engineer the L3 organization. However, clearing up the whole L3 data cache of 512 KB, even with thread level parallelism, substantially reduces bandwidth. Figure \ref{fig:LLC_Cov_1} shows the bandwidth of the LLC based covert channel is 1 kb/s, when the whole L3 is cleared in every iteration. 
The next level of optimization implemented to increase the channel bandwidth is to identify minimum number of addresses that is required to evict all the target addresses in the same LLC set with the constraint that these addresses reside  in separate LLC sets from the target set. Otherwise, it would create noise due to self eviction.  This approach requires only knowledge of the LLC and not L3. This bandwidth we achieved using this technique, is 70 kb/s for GPU-to-CPU channel (67 kb/s for CPU-to-GPU channel). Further optimization was achieved by carrying out the complete L3 reverse engineering and creating its eviction sets, determining the addresses that are in the same L3 set for precise eviction of the target addresses. This next fold of optimization increases the bandwidth to 120 kb/s (118 kb/s for CPU-to-GPU channel), which is a respectable bandwidth given that each GPU memory reference to the LLC must first be evicted from the L3. 
The error percentage of the precise L3 eviction came out to be as low as 2\% (6\% for CPU-to-GPU channel). We achieved a stable channel with a low error rate and high bandwidth through our optimization of precise L3 set eviction. However, the error rate is higher in the case of CPU-to-GPU channel. The primary reason is the increased usage of the custom timer during the first stage of handshake as well as during the bit communication that misinterprets the misses as hits. 

\begin{figure}[h]
\vspace{-3mm}
\centering
  \includegraphics[width=\linewidth]{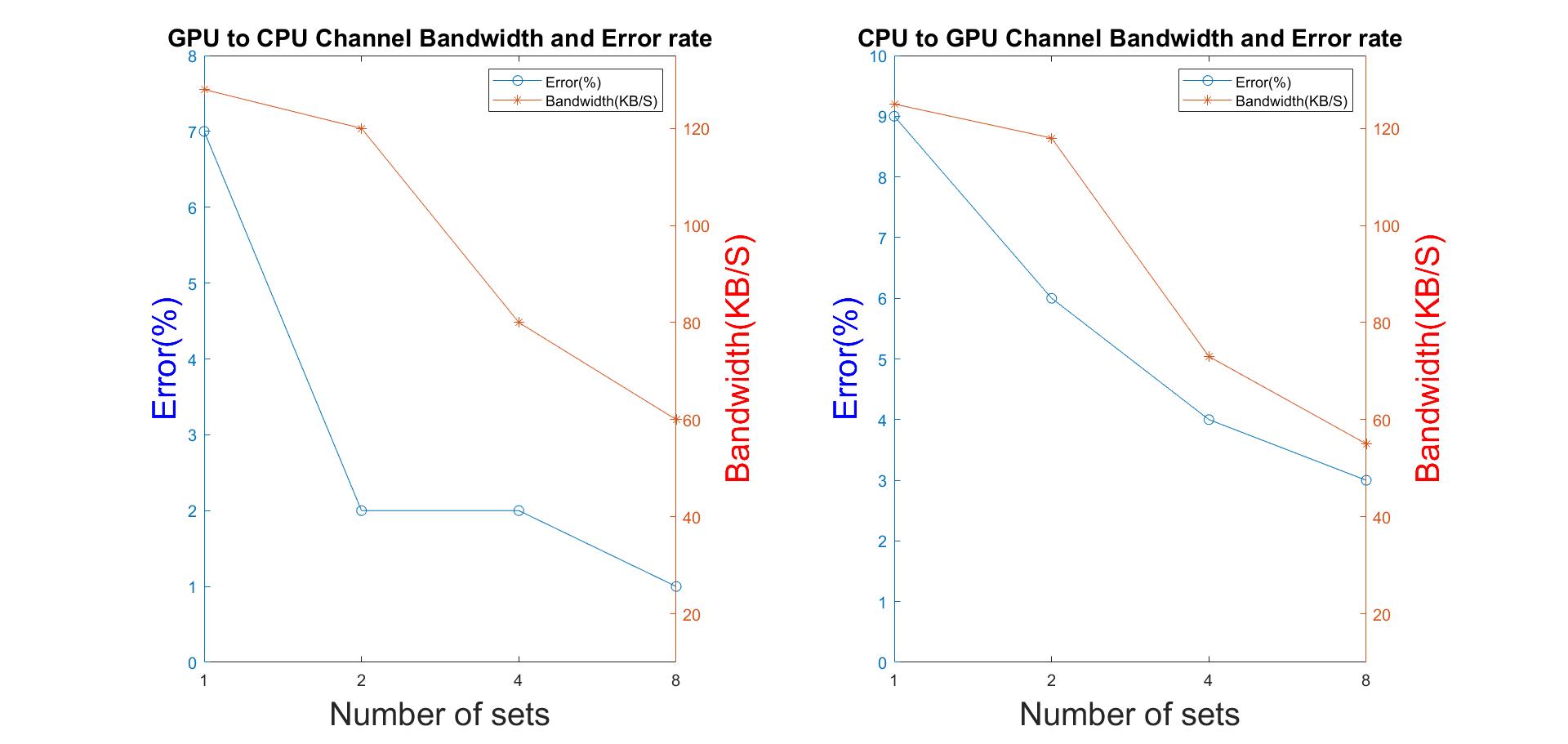}
\caption{\textbf{Error and BW with number of LLC sets}}
\vspace{-0.15in}
\label{fig:LLC_Cov_2}
\end{figure}

To reduce the error rate and increase channel resilience we used multiple LLC sets. Monitoring cache misses over multiple sets provides us with better resolution than using a single set for communication. However, the redundancy causes reduction in available bandwidth; potentially we could have used these multiple sets to communicate multiple bits in parallel. Figure \ref{fig:LLC_Cov_2} shows the bandwidth and error rate with respect to the increasing of number of LLC sets. When we are using only 1 set then the error rate is 7\% for GPU-to-CPU channel (9\% for the CPU-to-GPU channel), which reduces to 2\% as the number of sets doubled. For CPU-to-GPU channel that error rate reduces to 6\%. However, the bandwidth reduces by 6.25\% from 128 Kb/s to 120 Kb/s which is acceptable reduction given the error rate reduces by more than 71\%. The bandwidth reduces to 118 Kb/s from 125 Kb/s by doubling the cache set in the cases of CPU-to-GPU based channel. Increasing the number of sets does not provide any improvement on the error rate. However, the bandwidth reduces at a steady rate. In our attacks, we used 2 sets for all the 3 stages of attack resulting in using 6 LLC cache sets.

\cut{\sankha{The reverse cache based channel have similar bandwidth but with a higher error rate than the GPU to CPU cache based channel. The primary reason is the custom timer resolution. The usage of the custom timer in the reverse channel more than other channel due to which the resolution gets lost over the period of communication.}}

\begin{figure}[h]
\centering
  \includegraphics[width=0.7\linewidth]{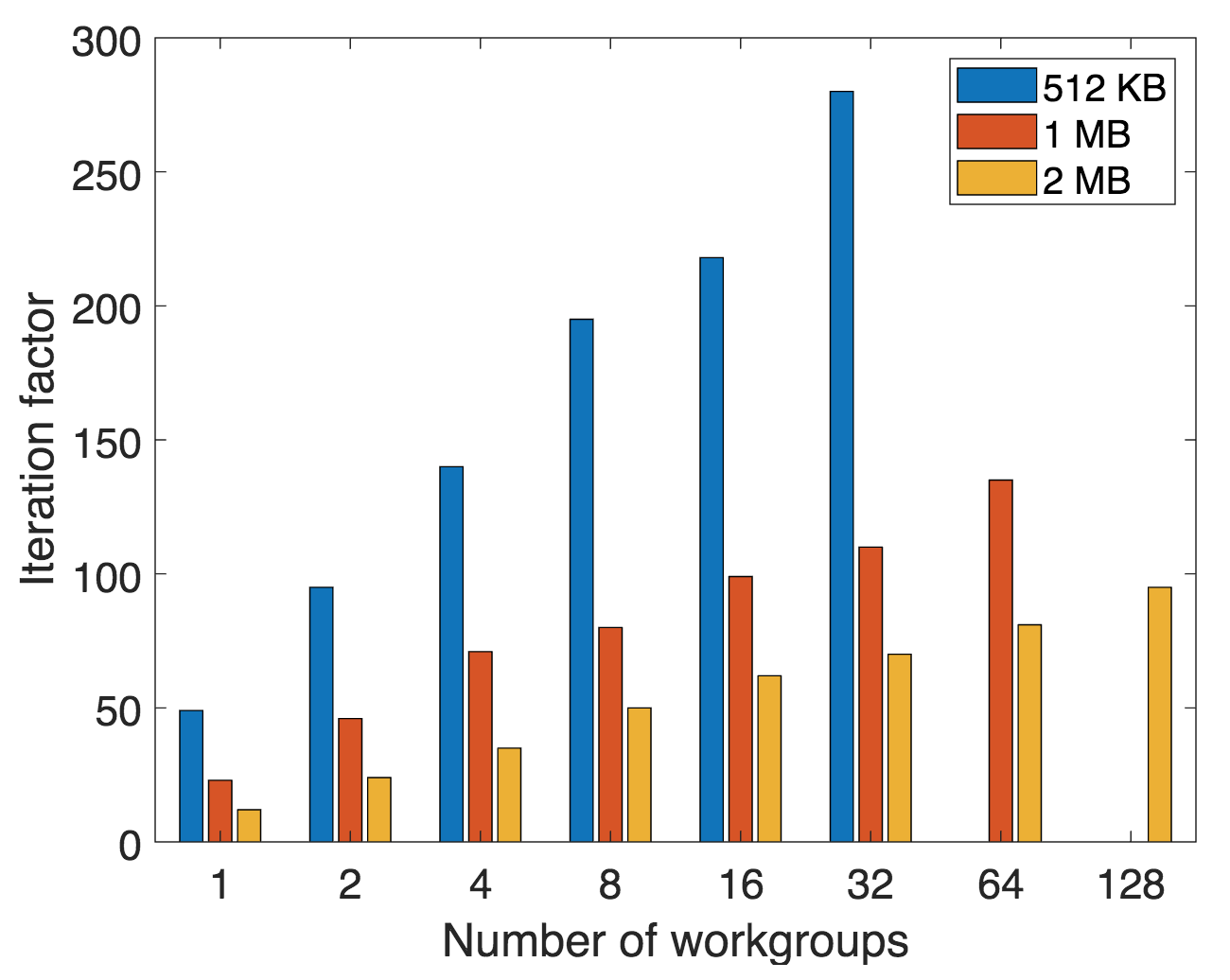}
\caption{\textbf{Iteration Factor for different buffer sizes}}
\vspace{-3mm}
\label{fig:freq_eq}
\end{figure}

\noindent\textbf{Contention-based Covert Channel:}
CPU and GPU access the LLC using asymetric pathways and computational models.  This impacts the success rate of the communication between the two asymmetric sides.  We introduce the concept of \textit{Iteration Factors} to match the rate of communication between the two sides  (as discussed in Section~\ref{sec:mem_request}). Figure \ref{fig:freq_eq} shows the optimal iteration factor: keeping the CPU buffer size constant, as the GPU buffer size increases, the factor reduces correspondingly to enable overlap between the two sides.

\begin{figure}[h]
\vspace{-3mm}
\centering
  \includegraphics[width=1\linewidth, height=0.7\linewidth]{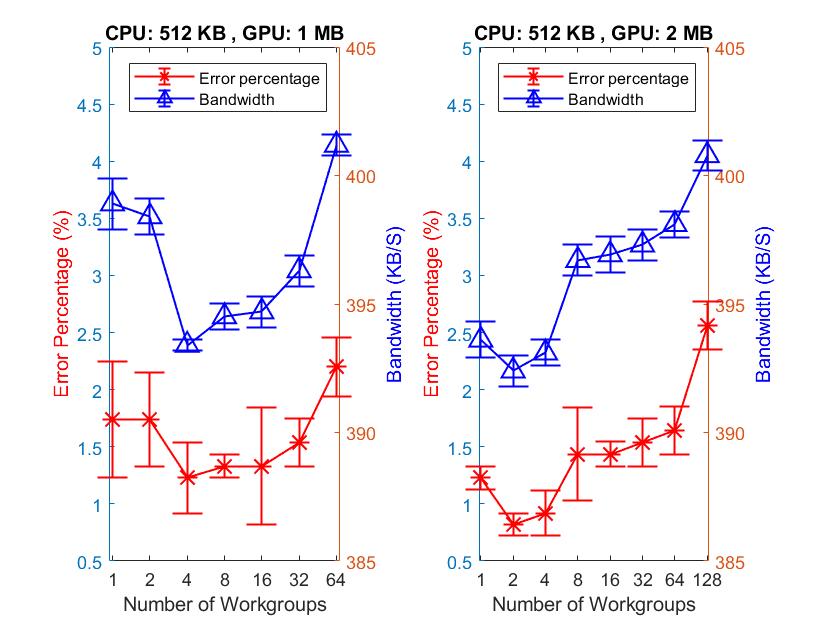}
\caption{\textbf{Bandwidth and error for bus-based channel}}
\label{fig:busCov_2}
\vspace{-3mm}
\end{figure}

As discussed in Section~\ref{sec:mem_request}, in our contention based covert channel, buffer size on both CPU and GPU side and the number of work-groups that access to the GPU buffer, affect the contention pattern and consequently the channel bandwidth and error rate. We perform a search on the parameter space to obtain a channel with low acceptable error rate and high bandwidth. Figure \ref{fig:busCov_2} shows the evaluation results of the contention based covert channel. The different graphs are for different GPU buffer size and a constant CPU buffer size of 512 KB. The GPU buffer size that we considered are 1 MB, and 2 MB.  Each result shows a confidence interval of 95\% over 1000 runs of the experiment. The bandwidth and the error rate are shown for different number of work-groups (in the X-axis). We obtained an error rate which is lower than 2\% for more than 90\% of the configuration space. The lowest error rate that we obtained is 0.82\% for CPU buffer size of 512 KB, GPU buffer size of 2 MB and number of work-groups of 2. We can observe that the bandwidth follows the pattern of the error rate (lower bandwidth for low error rate). 
 The bandwidth ranges from 390 kb/s to 402 kb/s. The bandwidth corresponding to the lowest error rate is 392.4 kb/s.


\section{Possible Mitigations}\label{sec:discussion}

We believe classes of defenses that have been developed against other microarchitectural covert channels can also potentially apply to cross component attacks on heterogeneous systems. These solutions include:
(1) Static or dynamic partitioning of resources~\cite{Kong-09,Wang_ISCA'07,domnitser-12,liu-16, QoS-2007}, specifically the LLC. These partitioning schemes can be extended to support different processors in heterogeneous systems.  If the Spy and Trojan use different partitions of the cache, they are not able to replace each other's cache lines; and (2) Eliminating the contention among processes by traffic control in memory controllers~\cite{Wang_hpca14,Shafiee_micro15}, such that memory requests from each processor are grouped into the same queue and possibly access the same memory bank/port. Prior work~\cite{Fang-2020} demonstrated that an efficient memory scheduling strategy and isolating the CPU memory requests from the GPU memory requests will improve the system performance, since memory requests from the GPU seriously interfere with the CPU memory access performance. Such isolation can be applied to the ring bus connecting the CPU and GPU to the LLC (with LLC partitioning in place). Other solutions such as adding noise to the timer may also apply~\cite{martin-12}.  However, we build our customized timer using hardware resource (shared memory) available on GPU, so disabling the timing infrastructure in our attack is not straightforward.  
 
 \section{Related Work}\label{sec:related}
Microarchitectural covert-channel and side-channel attacks have been widely studied on different resources on CPUs including the L1 cache~\cite{percival-05, brumley-2009, Tromer-2009} and shared LLC in multi-core CPUs~\cite{mehmet-2016, liu-15, Gruss-2015, ARMMagedon, Yarom-2014, Zhang-2014}. Some works exploit cache coherency protocols to develop timing channels on multi-core CPUs~\cite{fan_hpca2018, Yan-2019} or multi-CPU systems~\cite{Gorka-2016}. 

Some recent work demonstrates that GPUs are also vulnerable to microarchitectural covert and side-channel attacks. These work has been proposed on discrete GPUs with a dedicated memory. Jiang et al.~\cite{jiang-2016,jiang-2017} present architectural timing attacks from the CPU to the GPU. The attack triggers an AES computation on the GPU, and times it showing that there exists correlation the measured time (which varies due to key dependent memory coalescing behavior) and the last round key in AES encryption. 
Naghibijouybari et al.~\cite{Hoda-MICRO} construct several types of covert channels on different resources within a GPU.    
Naghibijouybari et al. also demonstrate a series of end-to-end GPU side channel attacks covering the different threat scenarios on both graphics and computational stacks, as well as across them  ~\cite{Hoda-CCS}.  They implements website fingerprinting through GPU memory utilization API or GPU performance counters, track user activities as they interact with a website or type characters on a keyboard. In addition, they develop a neural network model extraction attack, demonstrating that these attacks are also dangerous on the cloud. 
On the defense side, Xu et al.~\cite{GPUGUARD} proposed a GPU-specific intra-SM partitioning scheme to isolate contention between victim and spy and eliminate contention based channels after detection.
 
All of these microarchitectural attacks and defenses have been proposed on a single processor (CPU or discrete GPU).In this paper, for the first time, we develop microarchitectural covert channels in more widely used integrated CPU-GPU systems. The integrated GPU is available through APIs such as WebGL~\cite{webgl} even for remote Javascript programs making this threat vector extremely dangerous.  There have been a limited number of attacks on heterogeneous systems (but not timing attacks):  Weissman et al.~\cite{Jackhammer-2020} study rowhammer attacks on heterogeneous FPGA-CPU platforms.  Frigo et al.~\cite{Glitch-2018} use WebGL timing APIs to implement GPU accelerated rowhammer attack on memory in integrated CPU-GPU systems in mobile SOCs. They use WebGL timer to find contiguous area of physical memory to conduct the rowhammer. We investigate a different threat model, microarchitectural covert channels, showing for the first time that these attacks can apply across components in a heterogeneous system.

\section{Concluding Remarks}\label{sec:conclusion}

To the best of our knowledge, we present the first microarchitectural covert channel attacks that span two different components in an SoC with an asymmetric view of the resource.  Specifically, each component has a different view of the shared resource that they use to create contention.  Beyond the extra complexity of reverse engineering two different pathways to the shared resource, this also introduces additional novel difficulties that arise due to this asymmetry.  For example, the LLC is inclusive on the CPU side, but non-inclusive on the GPU side.  Moreover, the indexing of the GPU cache hierarchy is different from that of the LLC; as we create conflict sets to overflow the L3 on the GPU, we run the risk of creating self-interference with other sets on the LLC.  We also needed to calibrate the communication loops to improve the bandwidth given the asymmetric pathways to access the channel.   

Although we demonstrated one instance of cross-component attacks (specifically, an integrated GPU and CPU) in heterogeneous systems, the threat model can be extended to include any other accelerator or components, sharing resources with CPUs. Having experience with these channels improves our understanding of the threats posed of microarchitectural attacks beyond a single component which is a threat model increasing in importance as we move increasingly towards heterogeneous computing platforms. We created two working channels: a Prime+Probe channel targeting the LLC, and a contention based channel exploiting contention on the shared access pathway to the LLC. Creating the channels required overcoming a set of challenges that we believe will be representative of those needed for cross-component attacks.  Both channels achieve high bandwidth and low error rate. 
\section{Acknowledgement}\label{sec:ack}
Pacific Northwest National Laboratory is operated by Battelle Memorial Institute for the U.S. Department of Energy under Contract No. DE-AC05-76RL01830. This work was supported by the U.S. Department of Energy, Office of Advanced Scientific Computing Research (ASCR) through the Center for Advanced Technology Evaluation (CENATE) project, Contract \#66150B. This work is also partially supported by National Science Foundation grant CNS-1619450.

\bibliographystyle{IEEEtranS}
\bibliography{refs}

\end{document}